\documentclass{article}

\usepackage{arxiv}
\usepackage{amsmath}
\usepackage[utf8]{inputenc} 
\usepackage[T1]{fontenc}    
\usepackage{hyperref}       
\usepackage{url}            
\usepackage{booktabs}       
\usepackage{amsfonts}       
\usepackage{nicefrac}       
\usepackage{microtype}      
\usepackage{lipsum}
\usepackage{multicol}
\usepackage[english]{babel}
\usepackage{amsthm}
\usepackage{graphicx}
\usepackage{subfig}
\usepackage{quiver}

\usepackage[capitalise,nameinlink,noabbrev]{cleveref}
\crefname{equation}{}{}

\theoremstyle{definition}

\theoremstyle{remark}

\title{Rotation curve of the Milky Way and the mass of Sagittarius A*}

\author{
 	J. Enrique H. Ramírez \\
 Department of Physics\\
 ABC Federal University\\
 Sao Paulo, Brazil\\
 \\
}

\begin{document}
\maketitle

\begin{abstract}
 In the present work we derive an analytical expression for the mass density of an object with spherical symmetry, whose corresponding potential allows obtaining a circular velocity around it that is in agreement with the observed rotation curve of galaxies. The rotation curve of our galaxy is analyzed, determining the properties of the central object, its radius and mass whose value obtained is very close to that reported in the literature.
\end{abstract}


\begin{multicols}{2}

\section{Introduction}
The problem of the rotation curve of galaxies has generated various theories and profile models (\cite{sofue2020rotation}), to explain the non-correspondence between the theory and the  observational data. The stars and galaxies when rotating around their center should show a decay in their rotation speed as they move away from their center. Many and different observations made show the opposite, as they move away from the center, their speeds increase and then maintain an almost constant value. The speed of rotation of the stars is much greater than what is allowed, which does not correspond to the estimated mass of the galaxy, calculated as the result of the masses of all the stars visible in it. The most accepted theory refers to the presence of a non-visible mass (dark matter), with the consideration of which it is possible to explain the non-decaying shape of the rotation curve. In this work we will consider the presence of this non-visible mass, whose density will be determined and which obeys a certain law.

\section{The model}
We will build a density $\rho$ that allows us to explain its properties as well as its behavior with the objects with which it interacts. We will deal, in this model, with a central object without rotation and with spherical symmetry.

The dimension $r$ of the object is obviously a magnitude to consider in the expression of the density that we want to build. As daily experience shows us, the density in large objects, such as planets and stars, is not constant, it depends on which point in the interior of the object we refer to. If we are close to the surface of the object, its density will be less than if we refer to any point very close to its center. This leads us to consider that the density is inversely proportional to the radius of the object, $\rho \propto 1/r^{n}$, where $n$ is an integer to be determined. As we are dealing with problems related to gravitation, we will also consider a magnitude that characterizes the interaction of the object of study with other objects around it, the universal gravitational constant $G$. The other magnitude to consider, which characterizes the internal properties of the object of study, is its escape velocity, $v_{es}$.

Thus we can express the density as,

\begin{eqnarray}\label{densitydefinition1}
\rho(r) = k\, G^{x}\, v_{es}^{y}\, r^{n},
\end{eqnarray}

where $k$ is a dimensionless proportionality constant and $x$, $y$ and $n$ numbers to be determined. A dimensional analysis gives us that $x=-1$, $y = 2$ and $n =2$, so,

\begin{eqnarray}\label{densitydefinition2}
	\rho(r) = k\frac{v_{es}^{2}}{G\, r^2}.
\end{eqnarray}

As we known, the escape velocity of an object of mass $M$ and radius $a$ is given by, $v_{es}^{2} = 2\, M\, G/a$, thus, we can write the above equation as

\begin{eqnarray}\label{densitydefinition4}
	\rho(r) = k\frac{2\, M}{a\, r^2}.
\end{eqnarray}

We will determine the proportionality constant $k$, knowing that the mass contained in the volume $4/3\,\pi\, a^3$ is $M$. Thus, $k = 1/8\,\pi$. In this way we can finally express the density as

\begin{eqnarray}\label{densitydefinition5}
	\rho(r) = \frac{1}{3}\, \rho_{obj}\left(\frac{a}{r}\right)^2,
\end{eqnarray}

 being $\rho_{obj} = 3M/4\,\pi a^3$.

The mass corresponding to the distribution given by \ref{densitydefinition5} is

\begin{eqnarray}\label{masa1}
	m(r) = M\,\frac{r}{a}.
\end{eqnarray}

Note that for small distances from the center of the distribution, the density given by \ref{densitydefinition5} is large but the mass given by \ref{masa1} is small, but for large distances the opposite is true.

It should be noted that the above density obtained analytically is it very similar to that used by \cite{navarro1997universal}, obtained through high-resolution N-body simulations, to study the equilibrium density profiles of dark matter in halos.

\section{Potential}
Now we are going to consider the potential $\phi$ generated by this mass distribution. We will consider that this density is not limited to the border $a$ of the object, but that it extends beyond it. That is, the space that surrounds our object of study is not empty. In this model we only consider the mass distribution given by \ref{densitydefinition5} and the vacuum, and that the object is not rotating.

Solving the Poisson's equation for this mass distribution, we found that

\begin{eqnarray}
	\phi(r) = \frac{M\,G}{a}\, \left[-\ln(r/a) - c_{1}\,a/r + c_{2}\right],
\end{eqnarray}

where by the boundary conditions $\phi(a) = -G\,M/a$ and $\phi\prime (a) = -G\,M/a^2$ we found that $c_{1} = 2$ and $c_{2} = 1$, thus the potential will be given by

\begin{eqnarray}\label{potential1}
	\phi(r) = \frac{M\,G}{a}\, \left[-\ln(r/a) - 2\,a/r + 1\right].
\end{eqnarray}

\section{Rotation Velocity}

Now we can determine the rotation velocity $v_{rot}$ of some particle of mass $m$ around our central object, considering that the space around it is not empty but rather filled with an unknown substance whose density obeys the distribution given by \ref{densitydefinition5} . 

\begin{figure}[h]
	\centering
	\includegraphics[width=0.7\linewidth, height=0.3\textheight]{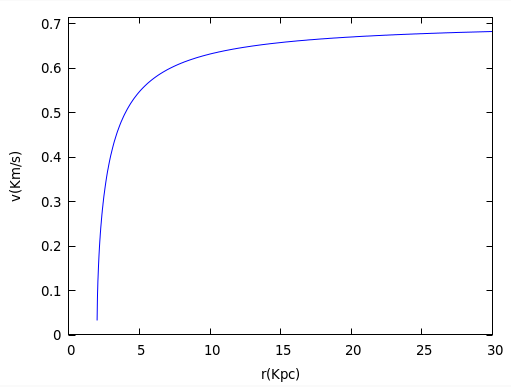}
	\caption[CURTA]{The curve is valid for $r\geq 2 a$ . As shown in the figure, the circular velocity does not decrease and approaches to $\sqrt{G\,M/a}$ as $r\to \infty$. Here we set $a = 1$ and $GM = 0.5$ for illustrative porpouses.}
	\label{figura1}
\end{figure}

The  rotation velocity $v$ can be determined as,

\begin{eqnarray}
	\frac{v^{2}}{r} = -\frac{d \phi}{d r},
\end{eqnarray}

thus we find that,

\begin{eqnarray}\label{cuvarot1}
	v = \sqrt{\frac{G\, M}{a}\left(1-\frac{2\, a}{r}\right)},
\end{eqnarray}

whose graph is shown in (Fig.\ref*{figura1}). 
\end{multicols}
\begin{multicols}{2}
This simple but extremely useful expression will allow us to determine the main characteristics of the celtral object, such as its mass $M$ and dimension $a$, as we will see later.

For long distances ($r\gg 2 a$),

\begin{eqnarray}
	v \approx \sqrt{\frac{G\, M}{a}}\left(1-\frac{\, a}{r}\right).
\end{eqnarray}

\section{Fitting the Galaxy Rotation Curve and Mass of the Galactic Center}

In this section we will show the rotation curves of the Milky Way and how to find the fundamental characteristics of the central object, its mass($M$) and radius($a$). The observational data used in this work were taken from \cite{sofue1999central}.

We can rewrite equation \ref{cuvarot1} in a convenient way

\begin{eqnarray}\label{linearizaciondecurva}
Y = B - A\, X
\end{eqnarray}

where $Y = v^2$, $A = 2 G\,M$, $B = G\,M/a$ and $X = 1/r$.
  Determining, from Eq. \ref{linearizaciondecurva} , the constants $A$ and $B$ using the observational data, we were able to find the mass $M$ and the radius $a$ of the central object.

\begin{figure}[h]
	\centering
	\includegraphics[width=1.1\linewidth, height=0.5\textheight]{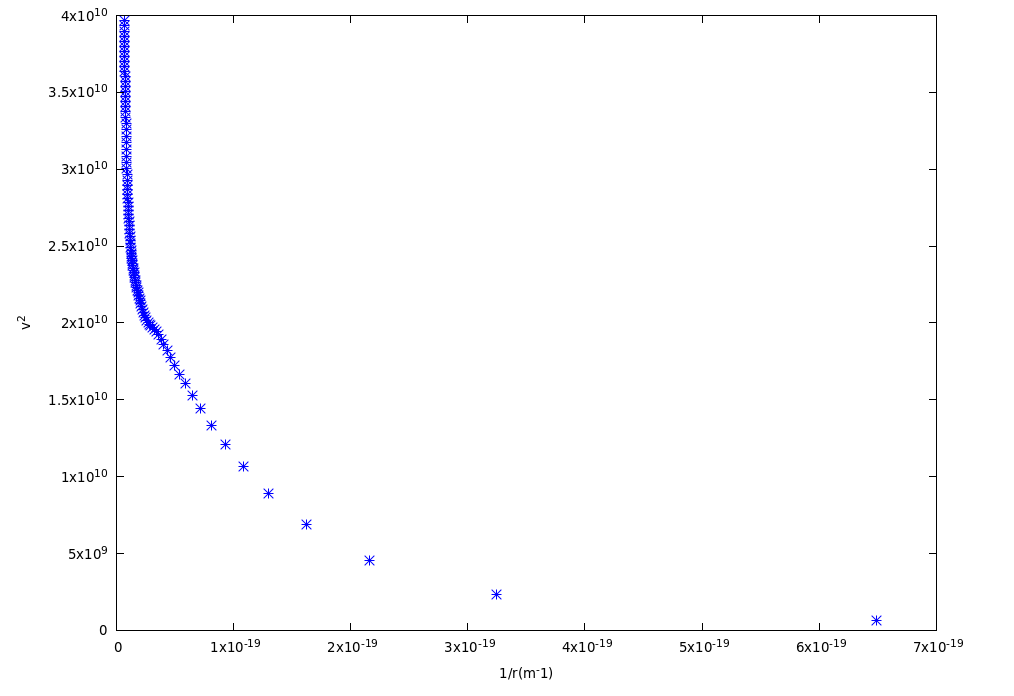}
	\caption[curto]{The figure shows the graph of $1/r$ vs $v^2$ according to the data show in Fig. \ref{tabla1}. Because the mass varies with $r$, it does not show a linear dependence according to the proposed model.}
	\label{lineardataexample1}
\end{figure}

In the case of our galaxy, the curve is shown in the Fig.\ref{lineardataexample1}. This does not show the linear behavior predicted by the model,  because it  considers that the central object has a fixed mass $M$ and radius $a$. In the case of galaxies, this mass varies, due, on the one hand, to the large number of objects between the central object and the object that orbits and due to the non-visible mass, whose density distribution is given by\ref{densitydefinition5} . However, we note that certain groups of data have a linear behavior. For this reason, we split the data into five groups, using the high degree of linearity in each of them as a criterion. The first group from 1-4, the second from 5-21, the third from 22-41, the fourth from 42-68 and the last from 69-99. The equations of the lines, according to equation \ref{linearizaciondecurva}, are:

 $Y_{1} = -1.1177\times 10^{25} +7.3641\times 10^{9}$,\\
 $Y_{2} = -1.1543\times 10^{26} +2.3121\times 10^{10}$,\\
 $Y_{3} = -2.3857\times 10^{26} +2.643\times 10^{10}$, \\
 $Y_{4} = -9.3890\times 10^{26} +3.67\times 10^{10}$ and\\
 $Y_{5} = -3.7650\times 10^{27} +6.3970\times 10^{10}$, respectively. 
 
 Through them we were able to determine the corresponding masses, being them $M_{1} = 8.44\times^{34} \,kg$,$M_{2}= 8.72\times^{35} \,kg$,$M_{3}= 1.80\times^{36} \,kg$ and $M_{4}= 7.09\times^{36} \,kg$, $M_{5}= 2.84\times^{37} \,kg$, whose average is $M_{Sgr A^*}= 7.65\times 10^{36}\,kg$, value very close to that reported in \cite{abuter2018detection} (3.85 million solar masses). 

\begin{figure}[h]
	\centering
	\includegraphics[width=1.1\linewidth, height=0.5\textheight]{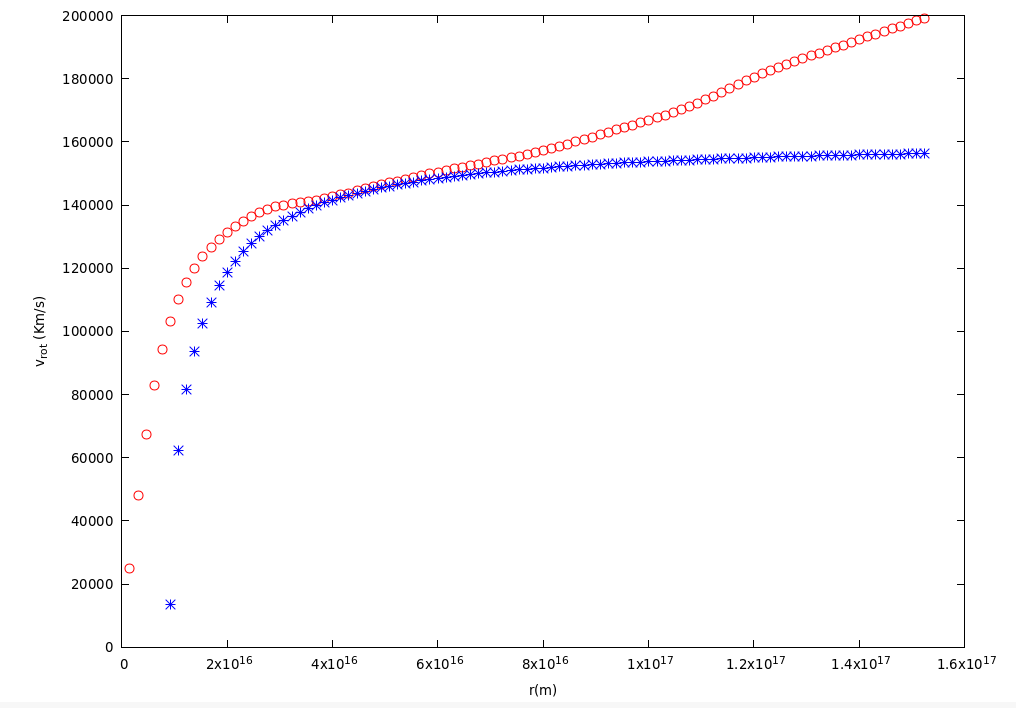}
	\caption[curto]{The figure shows the rotation curve obtained from the observational data (red circles) and the best fit according to our model (blue asterisks). $M_{3} = 1.80\times 10^{36}\,Kg $ and $a = 4.59\times 10^{15}\, m$ were used.}
	\label{curvarotacion2}
\end{figure}

The corresponding radius are: $a_{1} = 7.55\times 10^{14} \,m $ , $a_{2} = 2.51\times 10^{15} \,m $ , $a_{3} = 4.59\times 10^{15} \,m $ , $a_{4} = 1.27\times 10^{16} \,m $ , $a_{5} = 2.94\times 10^{16} \,m $ whose mean value is $a_{Sgr A^*} = 9.98\times 10^{15} \,m$ . The observational and fitted curves are shown in \ref{curvarotacion2} .

$M_{Srg A^*}$ and $a_{Srg A^*}$ are average values. The best approximation for these is privided by the two objects closest to the galactic center that orbit it. On the other hand, the best approximation to the mass  and dimension of the galaxy ($M_{galaxy}$ and $R_{galaxy}$) come from the two most distant objects that orbit it.
 
\section{Angular Momentum Conservation}

Since the Lagrangian of the system does not depend explicitly on $\theta$ the angular momentum $L$ is conserved. It happens that after a certain distance the speed of rotation remains almost constant, but not the position of the orbiting object, which increases. It may be asked on account of what angular momentum is conserved. Let's see,

\begin{eqnarray}
L = m\,v\, d,
\end{eqnarray}

being $m$ the mass of the star, $v$ its speed of rotation and $d$ its distance from the galactic center. Since $L$ is constant then, the mass must vary according to the equation

\begin{eqnarray}\label{variacionmasa}
m(r) = \frac{m_{0}\,v_{0}\, r_{0}}{r\,\sqrt{\frac{G M}{a}\left(1-2\,a/r\right)}}.
\end{eqnarray} 

As the star moves away from the center of rotation, it loses mass and with it, energy. The mass loss mechanism occurs by the well-known nuclear process that occurs in stars, in addition to the one exposed by conservation of angular momentum.

\section{Conclusion}	

The data obtained by the model was possible assuming the presence of a non-visible distribution of matter, outside the limits of the central object. The Eq. \ref{cuvarot1}  is in correspondence with the observational data as it does not present a decay and it is valid for spiral galaxies or others whose dependence $1/r$ vs $v^2$ is given by the Eq. \ref{linearizaciondecurva} . Since the speed of rotation of the stars depends on the mass of the galaxy contained in a radius $r$, several rotation curves are obtained, each of which will reproduce the observational data in a certain range. A more realistic model should include the rotation of the central object, this would allow a better fit of the model with the observational data.

	\bibliographystyle{acm}  

	\bibliography{references.bib}

\end{multicols}

\begin{figure}[h]
	\centering
	\includegraphics[width=0.8\linewidth, height=0.93\textheight]{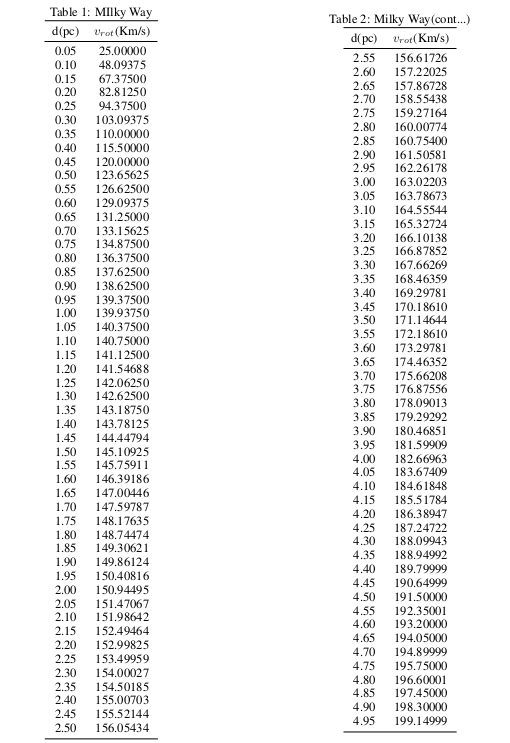}
	\caption[curto]{Rotation curve of the Milky Way}
	\label{tabla1}
\end{figure}

\end{document}